\documentclass[12pt]{article}
\usepackage{graphicx,amssymb,epsfig,epsf} 
 \usepackage{ulem}
\usepackage[usenames]{color}
\usepackage{rotating}
\usepackage{epsfig}
\usepackage{amsmath}
\usepackage{axodraw4j}
\usepackage{pstricks}
\usepackage{color}

\def\bt{\begin{table}}
\def\et{\end{table}}
\def\bc{\begin{center}}
\def\ec{\end{center}}
\def\bi{\begin{itemize}}
\def\ei{\end{itemize}}
\def\bea{\begin{eqnarray}}
\def\eea{\end{eqnarray}}
\def\beas{\begin{eqnarray*}}
\def\eeas{\end{eqnarray*}}

\def\beq{\begin{equation}}
\def\eeq{\end{equation}}

\begin{document}
\begin{flushright}
   {\bf OSU-HEP-14-02}\\
\end{flushright}

\vskip 30pt

\begin{center}

{\large Constraining Bosonic Supersymmetry from Higgs results and 8 TeV ATLAS multi-jets plus missing energy data}\\
\vskip 20pt
{{Kirtiman Ghosh\footnote{kirti.gh@gmail.com}}}, {Durmus Karabacak\footnote{durmas@ostatemail.okstate.edu}} and S. Nandi\footnote{s.nandi@okstate.edu}   \\
\vskip 10pt
{Department of Physics and Oklahoma Center for High Energy Physics,\\
Oklahoma State University, Stillwater, OK 74078-3072, USA.}\\

\end{center}

\vskip 20pt
\abstract
{

The collider phenomenology of models with Universal Extra Dimensions (UED) is surprisingly similar to that of supersymmetric (SUSY) scenarios. For each level-1 bosonic (fermionic) Kaluza-Klein (KK) state, there is a fermionic (bosonic) analog in SUSY and thus UED scenarios are often known as bosonic supersymmetry. The minimal version of UED (mUED) gives rise to a quasi-degenerate particle spectrum at each KK-level and thus, can not explain the enhanced Higgs to diphoton decay rate hinted by the ATLAS collaboration of the Large Hadron Collider (LHC) experiment. 
However, in the non-minimal version of the UED (nmUED) model, the enhanced Higgs to diphoton decay rate can be easily explained via the suitable choice of boundary localized kinetic (BLK) terms for higher dimensional fermions and gauge bosons. BLK terms remove the degeneracy in the KK mass spectrum and thus, pair production of level-1 quarks and gluons at the LHC gives rise to hard jets, leptons and large missing energy in the final state. These final states are studied in details by the ATLAS and CMS collaborations in the context of SUSY scenarios. We find that the absence of any significant deviation of the data from the Standard Model (SM) prediction puts a lower bound of about 2.1 TeV on equal mass excited quarks and gluons.}

\section{Introduction}
The twin primary goals of Large Hadron Collider (LHC) are to understand the mechanism for electro-weak symmetry breaking (EWSB) as well as uncover any new dynamics that may be operative at the scale of a few TeVs. There is a plethora of TeV scale scenarios have been proposed in the literature during last four decades.  In this endeavour, lots of attention have been paid to the theories with one or more extra space-like dimensions.

Extra dimensional theories can be classified into several classes. Models of ADD \cite{add} or RS \cite{rs} have been proposed to circumvent the long-standing hierarchy problem. In this framework, gravity lives in $(4+D)$ dimensions and the SM particles are confined to a 3-brane (a $(3+1)$ dimensional space) embedded in the $(4+D)$ dimensional bulk. 
There are some interesting generalization of these models in which the SM  particles are confined to a $(3+n)$-brane ($3+n+1$ dimensional manifold) embedded in a $(4+D)$ dimensional bulk \cite{NPB550, PRD66}.  There are also models in which the
SM particles are confined to a $3$-brane which is ``fat'' i.e. it
has an extension in the $(4+D)$ dimensional bulk \cite{PLB482}.   

On the other hand, there are class of models where some or all of the Standard Model (SM) fields can access the full space-time manifold.  One such example is Universal Extra
Dimension (UED) model \cite{UED}. This model assumes that all particles can propagate in the flat extra dimensions. Depending on the number of extra dimensions UED models can be classified into different classes \cite{bsusy,kkmass,2ued}. In its minimal version (mUED) \cite{bsusy,kkmass}, there is only one extra dimension $y$ compactified on a 
circle of radius $R$ ($S_1$ symmetry) and the Lagrangian is invariant under the SM gauge group ($SU(3)_C\otimes SU(2)_W\otimes U(1)_Y$) in 5D. An additional $Z_2$ symmetry, which identifies $y$ to $-y$ is required to get zero mode chiral fermions at low energy. The $Z_2$ symmetry breaks the translational invariance along the $5$th dimension and generates two fixed points at $y=0$ and $y=\pi R$. 
The low energy 
effective Lagrangian contains infinite number of Kaluza-Klein (KK) excitations 
(identified by an integer number n, called the KK number) for all the fields which are present in the higher dimensional Lagrangian. The zero modes of the KK towers are generally identified with the SM particles. The collider phenomenology \cite{collmUED} of level-1 excitations of the SM gauge bosons and fermions of UED scenarios is quite similar to that of a supersymmetric (SUSY) scenario.  The particle spectrum of SUSY scenarios include fermionic (bosonic) superparteners for the SM gauge bosons (fermions). Therefore, for each bosonic (fermionic) level-1 KK-particle in UED, there is a fermionic (bosonic) analog in SUSY. As a result, UED models are often known as bosonic supersymmetry \cite{bsusy}.


UED is an effective theory in 4 space-time dimensions (4D). Therefore, one needs to take into account all operators that are allowed by the gauge symmetry of SM and Lorentz invariance. In view of this fact there are two extensions of the minimal UED model without extending its particle spectrum. One can add vector-like mass terms for the 5D in the bulk Lagrangian. These terms are consistent with 5D Lorentz symmetry as well as the gauge symmetry of the SM. Phenomenology of these terms have already been considered in the literature in some details \cite{sued}. The second possibility is to add kinetic terms for all the 5D fields at the orbifold fixed-points, i.e., the boundaries of the bulk and the brane. These terms are known as boundary localized kinetic (BLK) terms. It is important to note that the BLK terms are only consistent with 4D Lorentz symmetry as well as the gauge symmetry. The BLK terms are not a priory known quantities (since they are related to ultra-violet (UV) completion for such scenarios) and thus would serve as extra free parameters of the theory. In the case of mUED, all BLK terms are assumed to vanish at the cut-off scale and are radiatively generated at the low scale that ultimately contributes as corrections to the masses \cite{kkmass}. In this article we have investigated the phenomenology of the one universal extra-dimension scenario with  vector-like mass terms for the bulk fermions and BLK terms, known as non-minimal Universal Extra-Dimension (nmUED) scenario \cite{kong,nmUED}. 

The phenomenology of mUED scenario is determined in terms of only two parameters namely, the radius of compactification ($R^{-1}$) and the cut-off scale ($\Lambda$) of the theory. Moreover, mUED gives rise to a particle spectrum which is nearly degenerate at each non-zero KK-levels. Therefore, the collider phenomenology of mUED is similar to the phenomenology of a SUSY scenario with compressed mass spectrum. In the framework of minimal supersymmetric SM (MSSM), we can play with a large number of free parameters and generate a compressed SUSY spectrum. However, most of the other SUSY scenarios with theoretically well motivated SUSY breaking mechanism at the high scale, in general, gives rise to wide splitting between the super partners masses at the electroweak scale. The quasi-degeneracy of mUED is removed in nmUED via non-zero BLK terms. Therefore, nmUED gives rise to very interesting consequences at the collider experiments. In this article, we have studied the parameter space of nmUED scenario in the context of recent LHC Higgs data and ATLAS search for multijet + missing transverse energy signature. 

 The plan of the article is the following. We will give a brief
description of the model in
the next section. The phenomenology of the nmUED model in the context of Higgs results and direct collider searches will be discussed in section 3. We summarize in the last section.

\section{Model}

The minimal universal extra dimensions (mUED) is defined on $ M^{4} \times S^{1}/Z_{2} $ space time where the extra spatial coordinate (y) is compactified on a orbifold with compactification radius R. In mUED all the Standard Model particles
allowed to propagate into extra dimension. The compactification enables 5D fields to be decomposed into zero mode, identified as the SM field and a tower of massive particle states which are identified by a integer, $n$, known as Kaluza-Klein number. By means of $S^{1}/Z_{2}$ orbifolding with two fixed points ($y = \pm L$ or equivalently $y=[0,\pi R]$) chiral fermion structure of the SM is obtainable and unwanted zero modes (both in fermion and the gauge sector) can be avoided. The translational symmetry along the fifth dimension gives rise to the conservation of KK number $n$. In general there can be KK number violating interactions at the fixed points of the orbifold. However, symmetry of these fixed-point interaction under the interchange of the fixed points gives rise to another conserved quantity known as KK-parity ($\equiv (-1)^{n}$). 
There are several phenomenological consequences of KK-parity conservation. For example. KK-parity conservation ensures the stability of LKP and makes it a good candidate for cold dark matter (CDM) \cite{dark_ued5,dark_ued6}. Moreover, as a consequence of KK-parity, level-1 KK-particles can only be pair produced and decay into lighter level-1 KK-particles only. It is important to note that conservation of KK-parity can be violated if fixed point interactions are not symmetric under the interchange of the fixed points. The consequences of KK-parity violation in UED scenarios is similar to the $R$-parity violation in SUSY scenarios. 

One of the simplest extension of mUED, though with nontrivial consequences, is to write an extra fermion bilinear (bulk mass term), which is compatible with $ 5$D Lorentz invariance and the gauge symmetry, is called the split-UED (sUED) \cite{sued}. In sUED KK-parity can still be maintained assuming that coefficient of the fermion bulk mass is odd function of the extra dimension $y$. There are several consequences of fermion bulk mass terms. For example, in sUED, the zero mode fermion wavefunctions have a nontrivial profile in the extra dimension y. As a result, the KK-number violating overlap integral of two zero mode fermions and one KK-gauge boson wave functions depends on the bulk mass term and gives rise to a tree-level coupling between two zero mode fermions and one level-2 KK-gauge boson. These tree-level interactions are highly constrained from the four-Fermi interaction data and put stringent limits on the bilinear fermionic bulk mass terms. 
Another immediate consequence is that the typical masses of nth level KK particles, which is around $ \sim n R^{-1} $ in mUED, will have a dependence in this extra term. 
    
Another possible extension of mUED can be done by including brane localized kinetic (BLK) terms, which manifestly respects (4D) Lorentz invariance and gauge symmetry, to the 5D theory. This is a reasonable extension of mUED since mUED being a 5D theory (and hence nonrenormalizable) one should include all possible terms allowed by symmetries of the theory. In mUED, KK particle masses are determined by the compactification scale and particle spectrum is generically degenerate. This degeneracy can be removed slightly by the radiative corrections. In BLK terms extended UED type models one can obtain much larger splitting in the spectrum compared to one with radiative corrections and hence enriching the phenomenology of the model. The KK-parity can still be conserved provided that symmetric BLT terms are chosen at the orbifold fixed points.

In this work, the model we consider combines The collider phenomenology of models with Universal Extra Dimensions (UED) is surprisingly similar to that of supersymmetric (SUSY) scenariosThe collider phenomenology of models with Universal Extra Dimensions (UED) is surprisingly similar to that of supersymmetric (SUSY) scenariosThe collider phenomenology of models with Universal Extra Dimensions (UED) is surprisingly similar to that of supersymmetric (SUSY) scenariosThe collider phenomenology of models with Universal Extra Dimensions (UED) is surprisingly similar to that of supersymmetric (SUSY) scenariosThe collider phenomenology of models with Universal Extra Dimensions (UED) is surprisingly similar to that of supersymmetric (SUSY) scenariosThe collider phenomenology of models with Universal Extra Dimensions (UED) is surprisingly similar to that of supersymmetric (SUSY) scenariosThe collider phenomenology of models with Universal Extra Dimensions (UED) is surprisingly similar to that of supersymmetric (SUSY) scenariosThe collider phenomenology of models with Universal Extra Dimensions (UED) is surprisingly similar to that of supersymmetric (SUSY) scenarios split-UED with BLK terms \cite{kong}. 
The action for fermion and the gauge boson which have been compactified on a $S^{1}/Z_{2}$ with radius R (or equivalently in extra dimension $y = [-L,L]$) can be written as a sum of actions involving only bulk terms ($\mathcal{S}_{bulk}$) and brane localized terms ($\mathcal{S}_{bdry}$).
\begin{eqnarray}
\mathcal{S}_{bulk} &=& \int \mathrm{d}^4x \int_{-L}^{L}\mathrm{d}y \bigg[\sum_{\mathcal{A}}^{G,W,B} -\frac{1}{4} \mathcal{A}_{MN}\mathcal{A}^{MN} + \sum_{\Psi}^{Q,U,D,L,E} i\bar{\Psi} \overleftrightarrow{D}_{M} \Gamma^{M}\Psi - M_{\Psi}   \bar{\Psi}\Psi \bigg ] \nonumber\\
\mathcal{S}_{bdry} &=&  \int \mathrm{d}^4x \int_{-L}^{L}\mathrm{d}y \bigg (\sum_{\mathcal{A}}^{G,W,B} -\frac{r_{\mathcal{A}}}{4} \mathcal{A}_{\mu\nu}\mathcal{A}^{\mu\nu} + \sum_{\Psi = Q,L} i r_{\Psi} \bar{\Psi}_{L}D_{\mu}\gamma^{\mu}\Psi_{L} \nonumber\\
&&~~~~~~~~~~~~~~~~~~+\sum_{\Psi = U,D,E} i r_{\Psi} \bar{\Psi}_{R}D_{\mu}\gamma^{\mu}\Psi_{R} \bigg )\times [\delta(y-L)+\delta(y+L)].
\end{eqnarray}

where $\bar{\Psi} \overleftrightarrow{D}_{M} \Psi = \frac{1}{2}\{\bar{\Psi}(D_{M}\Psi) - (D_{M}\bar{\Psi})\Psi\} $, $\Gamma^{M}=(\gamma^{\mu},i\gamma^{5})$ is the gamma matrices in 5D, $r_{i}$ with $i = \mathcal{A},\Psi$ is the brane localized terms for gauge bosons and fermions and $M_{\Psi}$ is the universal fermion bulk mass term. We can write fermion bare mass term in 5D as $M_{\Psi} = \mu \theta(y)$ (notice that KK parity is intact since $M_{\Psi}(-y) = - M_{\Psi}(y)$) where $\theta(y) = 2H(y) - 1$ is the step function with $H(y)$ being Heaviside function and $\mu$ being the fermion universal bulk mass. 

The KK-decomposition of the 5D fermions, gauge-bosons and scalars, in the framework of nmUED, have already been studied in Ref.~\cite{kong}. After integrating out the compactified dimension, the effective 4-dimensional Lagrangian can be written in terms of the respective zero modes and the KK excitations. It is instructive to take a glance at the KK mode expansions of the 5D fermions and gauge bosons. Defining ${\cal C}_n(x)={\rm Cos}(k_nx)$ and ${\cal S}_n(x)={\rm Sin}(k_nx)$, the KK expansion for the 5D fermion is given by, 
\beq
\begin{split}
& \Psi(x,y) =  \left(\frac{\mu}{(1+2 r \mu)\exp(2 \mu L)-1}\right)^{\frac{1}{2}}e^{\mu|y|} \psi_{L}^{(0)}\\ & + \sum_{odd ~n} \left(L-\frac{{\cal C}_n(L){\cal S}_n(L)}{k_n}+2 r {\cal S}_n^2(L)\right)^{-\frac{1}{2}} \left[ {\cal S}_n(y)\psi_{L}^{(n)} + \left(-\frac{k_n}{m_{f_n}}{\cal C}_n(y)+\frac{\mu}{m_{f_n}}\theta(y){\cal S}_n(y)\right)\psi_{R}^{(n)}\right] \\ 
&+ \sum_{even ~n}\left(L-\frac{{\cal C}_n(L){\cal S}_n(L)}{k_n}\right)^{-\frac{1}{2}}\left[\left(\frac{k_n}{m_{f_n}}{\cal C}_n(y) + \frac{\mu}{m_{f_n}} \theta(y){\cal S}_n(y) \right)\psi_{L}^{(n)} + {\cal S}_n(y) \psi_{R}^{(n)} \right], 
\label{fermion_WF}
\end{split}
\eeq
where, the wave numbers, $k_{n}$ are determined from following transcendental equations:
\beq
\begin{split}
&k_{n} {\cal C}_n(L)  = (r_{\Psi}(m_{f_{n}})^{2} + \mu ) {\cal S}_n(L)~~\text{ for n : odd}\\
&r_{\Psi}k_n {\cal C}_n(L) = -(1 + r_{\Psi} \mu){\cal S}_n(L)~~~~~\text{for n : even,}
\end{split}
\eeq
and the mass of level-n KK fermion is  
$
m_{f_n}= \sqrt{k_n^{2}+\mu^{2}} . 
$
The KK expansion for the 5D gauge bosons can be written as,
\beq
\begin{split}
\mathcal{A}_{\mu}(x,y) &= \left(2L\left[1+\frac{r_{\mathcal{A}}}{L}\right]\right)^{-\frac{1}{2}} \mathcal{A}_{\mu}^{(0)}(x)\\& + \sum_j\left[\left(L+r_{\mathcal{A}} {\cal S}_{2j-1}^2(L)\right)^{-\frac{1}{2}}{\cal S}_{2j-1}(y)\mathcal{A}_{\mu}^{(2j-1)}
 + \left(L+r_{\mathcal{A}}{\cal C}_{2j}^2(L)\right)^{-\frac{1}{2}}{\cal C}_{2j}(y)\mathcal{A}_{\mu}^{(2j)} \right]. 
\label{gauge_WF}
\end{split}
\eeq
The masses of the KK-gauge bosons can be determined by solving the following transcendental equations
\beq
\begin{split}
&\cot(k_{(2j-1)} L) = r k_{(2j-1)}~~~  \text{for odd KK modes~}\\
&\tan(k_{(2j)} L)   = - r k_{(2j)}~~~~~~  \text{for even KK modes} \\
\label{bosonM}
\end{split}
\eeq

After discussing the KK-decomposition of different 5D fields, we would like to discuss the constraints on different parameters of this model. First of all, the KK-expansion of fermions in Eq.~\ref{fermion_WF} shows that $ \frac{r_{\Psi}}{L} > \frac{\exp^{-2\mu L}-1}{2\mu L}$ should be satisfied in order to avoid ghosts and/or tachyons in the fermion sector. Similarly, for the gauge boson sector, Eq.~\ref{gauge_WF} tells us that $r_A/L$ has to be larger than -1. The bounds arising from the low-energy observables are studied in details in Ref.~\cite{kong}. The KK-parity conserving interaction of the form $ \mathcal{L}_{002n} $ where $ n=1 - \infty $, allows even KK-level $ W $ bosons contribute to coupling constant in four-Fermi interaction, bound of which can be incorporated as bounds on S, T, U parameters. These parameters scale as $ M_{W^{2n}}^{-2} $ so that for a fixed compactification scale $ R^{-1} $, increasing $ r_{\mathcal{A}} $ narrows down the allowed range for $ \mu $ since KK masses of gauge bosons decreases with increasing $ r_{\mathcal{A}} $. Another restriction again comes from four-Fermi interactions. Electroweak precision tests put a lower bound on the suppression scale of the four-Fermi interactions in a certain parametrization. In particular $ g_{200} $, coupling for two SM fermions and second level Z boson, has a dependence on $ \mu $ (it also vanishes for $ \mu = 0 $) and no significant dependence on $ r_{\mathcal{A}} $ for a fixed value of $ \mu $. This, combined with the fact that second-level Z mass decreases with increasing $ r_{\mathcal{A}} $, excludes $ r_{\mathcal{A}} $ up to 0.5 L for  $ \mu L=$ -0.1 and fixed $ R^{-1}$. We would like to note that for values of $ \mu $ in the vicinity of 0, the electroweak precision will be insensitive because, for instance, for $ 0 > \mu L > -0.03$ and $ R^{-1}\approx 850 $ GeV, $ g_{200} $ will be small and $ M_{Z^{2n}} $ will be heavy enough to escape from the upper bound on four-Fermi interaction suppression scale. CMS collaboration also looked for production of new Z-like gauge boson in s-channel and its decay to dilepton, and put bound on the production cross-section times branching ratio. This analysis, in particular, was able to extend the exclusion limit on $ r_{\mathcal{A}} $ beyond 0.5 L for $ \mu L = -0.1$. Note that both electroweak precision test and the collider searches will be insensitive to small values of $ \mu $ since for these values $ g_{200} $ will be very small. We have chosen $ \mu L =-0.02$ throughout our analysis.


\section{Phenomenology}
In this section, we will discuss the phenomenology of nmUED in the context of the LHC Higgs data and multi-jets plus $E_T\!\!\!\!\!/~$ searches at 8 TeV center-of-mass energy. In addition to the SM parameters, the Higgs and collider phenomenology of nmUED model depends on the compactification scale ($L=\pi R/2$), boundary parameters for the gauge bosons ($r_A,~A\supset G,W,B$) and fermions ($r_\psi,~\psi\supset Q,U,D,L,E$), fermion bulk mass ($\mu$) and the cut-off scale of the model ($\Lambda$). In our analysis, we consider universal boundary parameters for all quark and lepton families denoted by, $r_F$. However, the boundary parameter for gluon ($r_g$) is assumed to be different from the boundary parameters for the boundary parameters for the $SU(2)$ and $U(1)$ gauge bosons ($r_g \ne r_W=r_B$).

\begin{figure}[t]
\begin{center}
\includegraphics[angle =-90, width=0.49\textwidth]{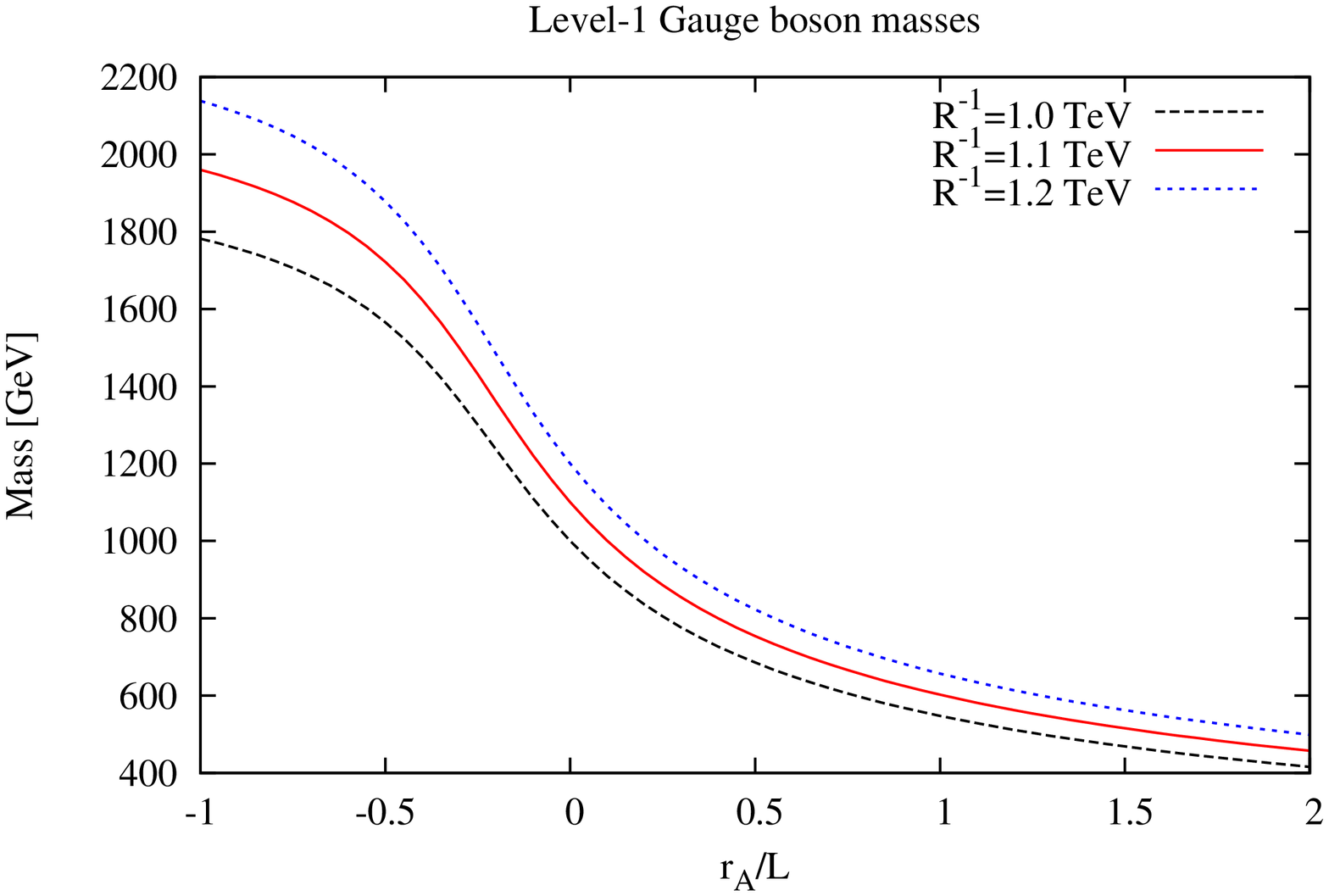}
\includegraphics[angle =-90, width=0.49\textwidth]{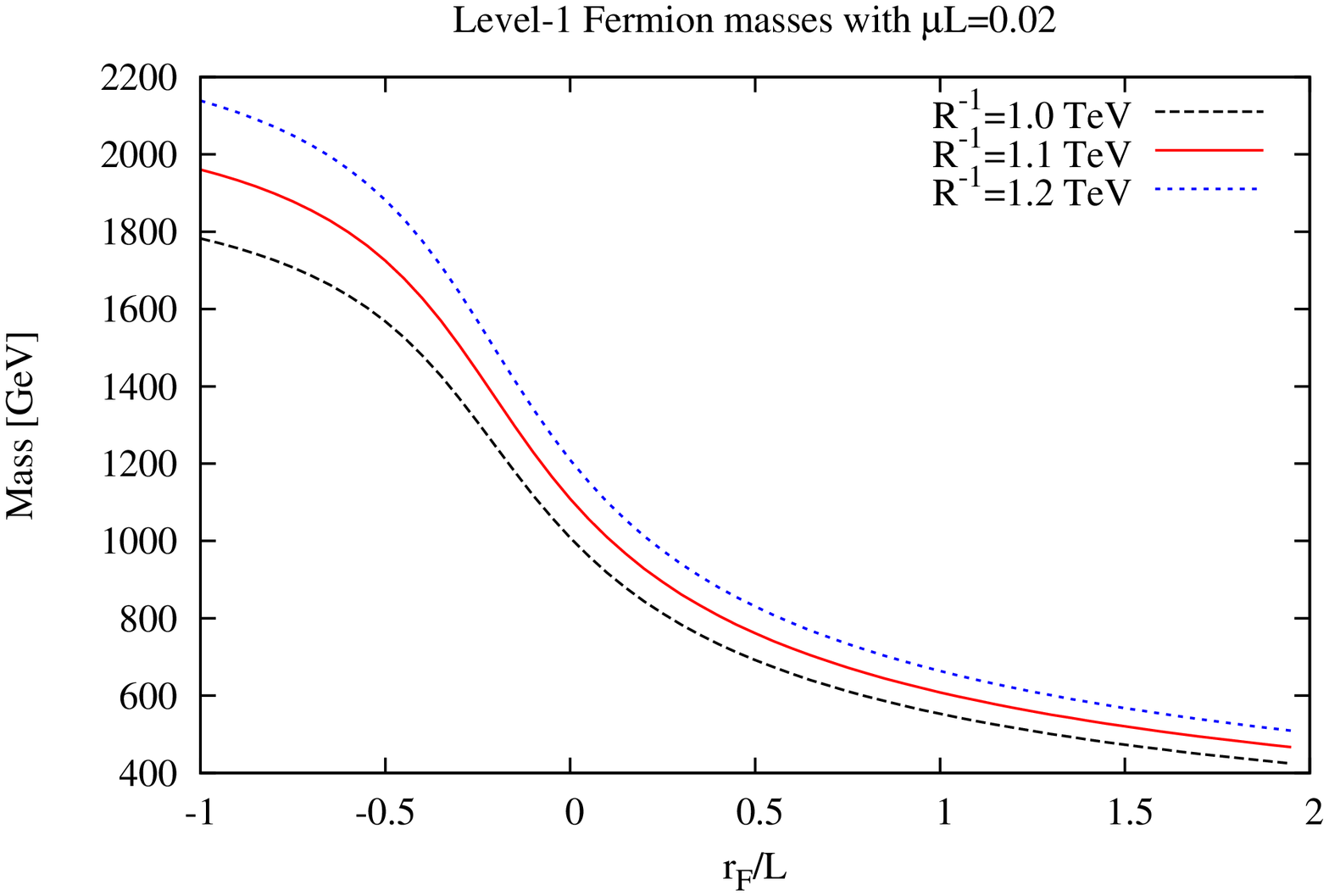}
\end{center}
\caption{Level-1 gauge boson (left panel) and fermion (right panel) as a function of $r/L$ for three different values of $R^{-1}$. For the level-1 fermion masses in the right panel, we consider $\mu=-0.02L$.}
\label{mass}
\end{figure}

As discussed in the previous section, the tree level KK gauge boson masses (see Eq.~\ref{bosonM}) are determined by $R^{-1}$ and $r_A$ and fermion masses get an additional contribution from $\mu$. In Fig.~\ref{mass}, we have presented level-1 gauge boson (left panel) and fermion masses (right panel) as a function of $r_A$ and $r_\psi$ respectively, for three different values of $R^{-1}$. It is important to note that in the presence of BLKT, the normalization-factors for the 5D gauge bosons in Eq.~\ref{gauge_WF} have nontrivial forms. In particular, the normalization factor for the zero mode field ($1/\sqrt{2L(1+r_A/L)}$, see Eq.~\ref{gauge_WF}) puts a theoretical lower bound on $r_A$ ($r_A/L>-1$) in order to avoid tachyonic zero mode. Therefore, in Fig.~\ref{mass} (left panel), we have varied $r_A/L$ between $-1$ to $2$. The normalization factor for the fermion zero mode in Eq.~\ref{fermion_WF} shows that in absence of fermion bulk mass terms, i.e., for $\mu=0$, the KK decomposition of 5D fermion fields gives rise to ghosts or tachyonic zero modes for $r_F<0$. Whereas, in presence of fermion bulk mass terms (for non-zero $\mu$), negative values of $r_F$ are theoretically allowed. However, it was already discussed in the previous section that the fermion bulk mass terms are highly constrained from the four-Fermi interactions and other low energy observables. Therefore, throughout in our analysis, we have considered $\mu=-0.02/L$ which is consistent with all the low energy observables \cite{kong} and allows us to vary $r_F/L$ over the negative values.  In Fig.~\ref{mass} (right panel), we have varied $r_F/L$ between $-1$ and $2$. Fig.~\ref{mass} shows that both level-1 fermion and gauge boson masses increase if we decrease the value of $r/L$. 
 
\subsection{Higgs Phenomenology}
The ATLAS \cite{atlas_higgs} and CMS \cite{cms_higgs} collaborations of the LHC have recently reported the discovery of a Higgs boson like scalar with mass $125$ GeV. Higgs mass is a free parameter in the UED models. Therefore, the discovery of the Higgs boson at $M_H \simeq 125$ GeV only determines the Higgs boson self-coupling $\lambda=M_H^2/2v^2 \simeq 0.129$ at the electroweak symmetry-breaking scale. There is no direct constraint on the other parameters of the theory from the discovery of 125 GeV Higgs boson mass. However, the running the Higgs boson self-coupling $\lambda(Q)$ gives rise to a indirect upper bound on the cut-off scale $\Lambda$ on the theory. In the SM, for $\lambda \sim 0.129$ at the electroweak scale,  there is a tendency for the running of $\lambda(Q)$ to be driven to smaller and smaller values, eventually becoming negative, if we increase the energy scale $Q$. As a result, the scalar potential becomes unbounded from below and the electroweak vacuum becomes unstable. In the SM, this happens at an energy scale in the ballpark of $10^{11}$ GeV \cite{SM_VSB}. In the framework of UED models, the presence of a whole set of Kaluza-Klein (KK) excitations of the SM particles results into faster running of $\lambda(Q)$ than the SM \cite{UED_VS1}. As a result, depending on the value of $R^{-1}$ in the TeV scale, the electroweak vacuum becomes unstable in the  ballpark of about 4-6$R^{-1}$ \cite{UED_VS2}. Therefore, through out in our analysis, we have considered $\Lambda=5 R^{-1}$.

One of the interesting consequences of a Higgs boson with mass about 125 GeV is that we can detect this Higgs in many different production and decay channels \cite{ref_higgs}. As a result, many distinct signal strengths, defined as production$\times$decay rates relative to the SM expectations, defines as $\mu_i=(\sigma\times {\rm BR})_i/(\sigma\times {\rm BR})^{\rm SM}_i$, have already been measured by the ATLAS and CMS collaborations. According to the latest communication by the ATLAS and CMS collaborations, almost\footnote{Some deviation in the gamma-gamma channel was reported by the ATLAS collaboration.} all of these measured signal strengths seem to coincide well with those expected in the SM. This poses constraints on various beyond the Standard Model (BSM) theories, in which Higgs signal strengths in different channels can differ substantially from those of the SM. In this work, we have studied constraints on UED scenarios in view of Higgs data. 

Particle spectrum of UED models contains towers of KK excitations of the SM particles. In the collider experiments, the absence of any new particle beyond the SM particle spectrum puts a lower bound of few hundred GeV on the masses of level-1 KK-particles. Therefore, the decay of Higgs with mass 125 GeV into a pair of KK-particles are kinematically forbidden. However, the loop induced couplings of the Higgs boson with a pair of SM photon and gluon get significant contributions from the KK-towers of the SM particles. In particular, KK-tower of the top quark contributes to the $Hgg$ coupling and both KK-tower of the top quark and $W$-boson contribute to the $H\gamma\gamma$ coupling. At the LHC, the dominant Higgs boson production channel is the gluon-gluon fusion channel and the cleanest final state for the discovery of the Higgs boson with mass in the  of $\sim 125$ GeV is the di-photon final state. Therefore, the measurement of Higgs signal strengths in different channels by the ATLAS and CMS collaborations imposes significant constraints on the parameters of UED models. 

 After LHC 7-8 TeV run, a large number of measurements of signal strengths now available (from both ATLAS and CMS collaboration) in different channels for different experimentally defined signal categories (based on combinations of cuts). Instead of checking the consistency of any BSM scenario with these huge number of measured signal strengths in different experimental categories by different collaborations, it is better to use the global fits of Higgs signal strengths defined by the Higgs production and decay modes only. In our analysis, we have used the fitted values of the Higgs signal strengths from Ref.~\cite{best-fit}. In principle, there are five different Higgs production mode namely, gluon-gluon fusion ($ggF$), vector boson fusion (VBF), associated production with a $W$ or $Z$-boson (${\rm VH}$), and associated production with a pair of top-quarks ($ttH$). However, experimental collaborations group these five production modes into just two effective modes namely, $ggF+ttH$ and $\rm{VBF+VH}$. In Ref.~\cite{best-fit}, the combined best-fit signal strengths ($\hat \mu^{ggF}$ and $\hat \mu^{\rm {VBF}}$) for these two production modes and different decay channels have been computed. In our analysis, we only consider $\gamma \gamma$ and $VV$ (where $V$ can be $W^\pm$ or $Z$-boson) decay modes and combined best-fit signal strengths in these two decay channel have been tabulated in Table~\ref{bfss}.

\begin{table}[h]
\begin{center}
\begin{tabular}{||c|c||c|c||}
\hline \hline
\multicolumn{2}{||c||}{$\gamma \gamma$ decay channel} & \multicolumn{2}{|c||}{$VV$ decay channel} \\
\hline
$\hat \mu^{ggF}$ & $\hat \mu^{\rm {VBF}}$ & $\hat \mu^{ggF}$ & $\hat \mu^{\rm {VBF}}$ \\\hline\hline
0.98$\pm$0.28 & 1.72$\pm$0.59 & 0.91$\pm$0.16 &1.01$\pm$0.49 \\\hline\hline

\end{tabular}
\end{center}
\caption{Combined best-fit Higgs signal strengths \cite{best-fit} for different Higgs production and decay modes.}

\label{bfss}
\end{table}

The effects of KK-fermions on the production and decay of Higgs bosons via the loop induced diagrams have already been studied in the literature by several authors \cite{KK-loop}. For the sake of completeness of this article, we are summarizing the relevant results in the following. In the SM, Higgs decays into a pair of gluons via a loop induced diagram involving only top quark. In framework of UED models the KK-excitations of the top quarks also contributes to the loop induced $H\to gg$ decay width. Whereas the loop diagrams that give rise to the decay of Higgs boson into a pair of photons involves both top quark and $W$-boson in the SM and KK-top quarks and KK $W$-bosons in the UED models. The partial decay widths of the Higgs boson into a pair of gluons and photons in the framework of UED models are given by,
\begin{equation}
\Gamma_{H\to gg}=\frac{\alpha_s^2}{8\pi^3}\frac{M_{H}^3}{v_{EW}^2}\left|J_{t}(M_H^2)\right |^2,~~~\Gamma_{H\to \gamma\gamma}=\frac{\alpha^2 G_F}{8\sqrt 2 \pi^3}M_{H}^3\left|\frac{4}{3}J_{t}(M_H^2) + J_{W}(M_{H}^2)\right|^2,
\end{equation}    
where, $\alpha=1/127$ is the fine-structure constant, $\alpha_s$ is the QCD coupling strength, $v_{EW}=246$ GeV  is the Higgs vacuum expectation value, $G_F$ is the Fermi constant and  $J_{t}(x),~ J_{W}(x)$ are the loop functions defined as,
\begin{eqnarray}
J_t(x)=&-&2\frac{m_t^2}{x}+\frac{m_t^2}{x}\left(1-4\frac{m_t^2}{x}\right)I\left(\frac{m_t^2}{x}\right)\nonumber\\
&+&\sum_n \left(\frac{m_t}{m_{t^{(n)}}}\right)\left[-2\frac{m_{t^{(n)}}^2}{x}+\frac{m_{t^{(n)}}^2}{x}\left(1-4\frac{m_{t^{(n)}}^2}{x}\right)I\left(\frac{m_{t^{(n)}}^2}{x}\right)\right],\nonumber\\
J_W(x)=&&\frac{1}{2}+3\frac{m_W^2}{x}-\left[\frac{m_W^2}{x}\left(3-6\frac{m_W^2}{x}\right)\right]I\left(\frac{m_W^2}{x}\right)\nonumber\\
&+&\sum_n \left(\frac{m_t}{m_{W^{(n)}}}\right)\left\{ \frac{1}{2}+4\frac{m_W^2}{x}-\left[\frac{m_W^2}{x}\left(4-8\frac{m_W^2}{x}\right)-\frac{m_{W^{(n)}}^2}{x}\right]I\left(\frac{m_{W^{(n)}}^2}{x}\right)\right\},\nonumber
\end{eqnarray}
 where,
\begin{equation}
I(\lambda)=\int_0^1\frac{dx}{x}\left[\frac{x(x-1)}{\lambda}+1-i\epsilon\right].
\label{loopfn}
\end{equation}
The first part of these loop functions (first line of $J_{t}(x)~{\rm and}~ J_{W}(x)$ in Eq.~\ref{loopfn}) corresponds to the contributions from the SM particles only. The UED contributions to the loop factors are given by a summation (second line of $J_{t}(x)~{\rm and}~ J_{W}(x)$ in Eq.~\ref{loopfn}) over all KK-excitations of the respective particle with mass below the cut-off scale ($\lambda=nR^{-1}$ with $n=5$) of the theory. We have used {\bf HDECAY} \cite{hdecay} package for computing the Higgs decay widths and branching ratios in the framework of the SM as well as UED\footnote{{\bf HDECAY} package only computes the SM decay widths of the Higgs boson. We have modified the {\bf HDECAY} package to include the contributions from the KK-particles and calculated the Higgs decay widths in the framework of UED models.} scenarios. The dominant Higgs production channel at the LHC is the gluon-gluon fusion which gets significant contributions from the KK-excitations of top quark. However, the parton level $gg\to H$ production cross-section is related to the $H\to gg$ decay width as
\begin{equation}
\hat \sigma (gg\to H)~=~\frac{\pi^2}{8 M_H}\Gamma_{h\to gg}(M_H)\delta(\hat s-M_H^2).
\end{equation}  
Therefore, $gg\to H$ production cross-section in the context of UED scenarios can be easily computed from $\Gamma_{h\to gg}$. All other relevant Higgs production channels at the LHC i.e., VBF, VH and ttH channel, involve only tree level Feynman diagrams at the leading Order (LO). Therefore, KK-particles do not contributes to the abovementioned production channels at the LO. In our analysis, we have computed the production cross-sections and branching ratios of the Higgs boson in different channels in the context of mUED and nmUED scenarios. Using these production cross-sections and branching ratios, we have calculated different Higgs signal strengths relative to the SM expectations ($\mu_i$) and compared these numbers with the combined best-fit values, listed in Table~\ref{bfss}, of the experimental results.

\begin{figure}[t]
\begin{center}
\includegraphics[angle =-90, width=1.0\textwidth]{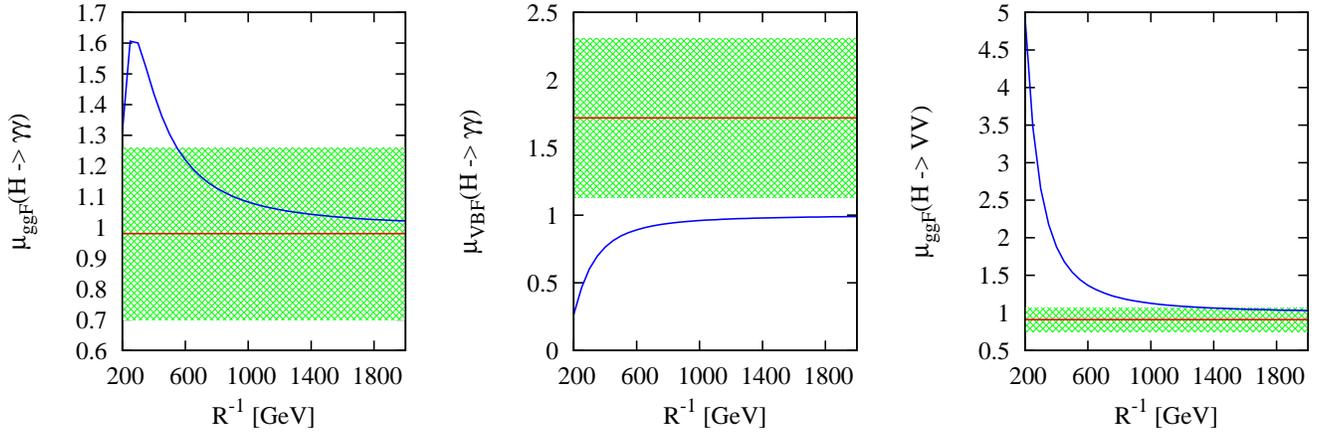}
\end{center}
\caption{Higgs signal strengths relative to the SM expectations: $\mu_{ggF}^{H\to \gamma\gamma}$ (left panel), $\mu_{VBF}^{H\to \gamma\gamma}$ (middle panel) and  $\mu_{ggF}^{H\to VV}$ (right panel), in the context of mUED scenario as a function of $R^{-1}$. The combined best-fit values (from Table~\ref{bfss}) of the abovementioned Higgs signal strengths are also presented.}
\label{mued_higgs}
\end{figure}

 The phenomenology of mUED scenario mainly depends only on value $R^{-1}$. In Fig.~\ref{mued_higgs}, we have presented $\mu_{ggF}^{H\to \gamma\gamma}$ (left panel), $\mu_{VBF}^{H\to \gamma\gamma}$ (middle panel) and  $\mu_{ggF}^{H\to VV}$ (right panel) as a function of $R^{-1}$. In Fig.~\ref{mued_higgs}, we have also presented the combined best-fit values (from Table~\ref{bfss}) of the abovementioned Higgs signal strengths. Before going into the discussion of Fig.~\ref{mued_higgs} it is important to note that for the loop induced Higgs couplings, the loops involving a particular tower KK-particles as well as the loop involving the corresponding SM particle  interfere constructively at the amplitude level. The $ggH$ loop involves only top quark in the SM and top quark as well as KK-excitations of the top quark in the UED scenarios. As a result, UED scenarios always gives rise to a enhanced $ggH$ coupling relative to the SM and hence, enhanced Higgs production cross-section in the gluon-gluon fusion channel. This feature of mUED model is manifested in the plot for $\mu_{ggF}^{H\to VV}$ (see Fig.~\ref{mued_higgs} right panel). Since the decay of Higgs boson into a pair of vector bosons talks place via tree level interactions only, in the computation of $\mu_{ggF}^{H\to VV}$, KK-particles contribute only in the production part and thus, give rise to an enhancement in the value of $\mu_{ggF}^{H\to VV}$. The enhancement is large for the smaller values of KK-top quarks masses i.e., smaller values of $R^{-1}$. Fig.~\ref{mued_higgs} (right panel) shows that mUED model is consistent with the combined best-fit value of $\mu_{ggF}^{H\to VV}$ only for $R^{-1}>1.2$ TeV. For Higgs to diphoton decay, both top quark and its KK-excitations as well as $W$-boson and its KK-excitations participate  in the loop. $W$-boson and top quark loop amplitudes interfere destructively and the SM (zero mode in mUED) $W$-boson (being lighter than the SM (zero mode) top quark) loop amplitude is larger than the top quark loop amplitude. However, mUED gives rise to nearly degenerate KK $W$-boson and KK-top quark masses at each non-zero KK-levels. For nearly equal mass of KK $W$-boson and KK-top quark,  KK-top quark loop amplitude is bigger than the KK $W$-boson loop amplitude. Therefore, for the zero mode, $W$-boson loop amplitude dominates and for the non-zero modes, KK-top quark loop amplitudes dominate. As a result, in the $\gamma\gamma H$ loop diagram, contributions from the non-zero KK particles interfere destructively with the contributions from the zero mode (SM) particles. Therefore, mUED always predicts suppression (relative to the SM expectation) in the Higgs to di-photon decay width and hence, in the Higgs to di-photon branching ratio. In Fig.~\ref{mued_higgs} (middle panel), we have plotted $\mu_{VBF}^{H\to \gamma\gamma}$ which is always below one for the mUED scenario. Whereas the the combined best-fit value for  $\mu_{VBF}^{H\to \gamma\gamma}$ is well above one (see Table~\ref{bfss}). Therefore, in the framework of mUED, it is not possible to satisfy the combined best-fit value for $\mu_{VBF}^{H\to \gamma\gamma}$ within its error bars for any values of $R^{-1}$.  

\begin{figure}[t] 
\begin{center}
\includegraphics[angle =-90, width=0.49\textwidth]{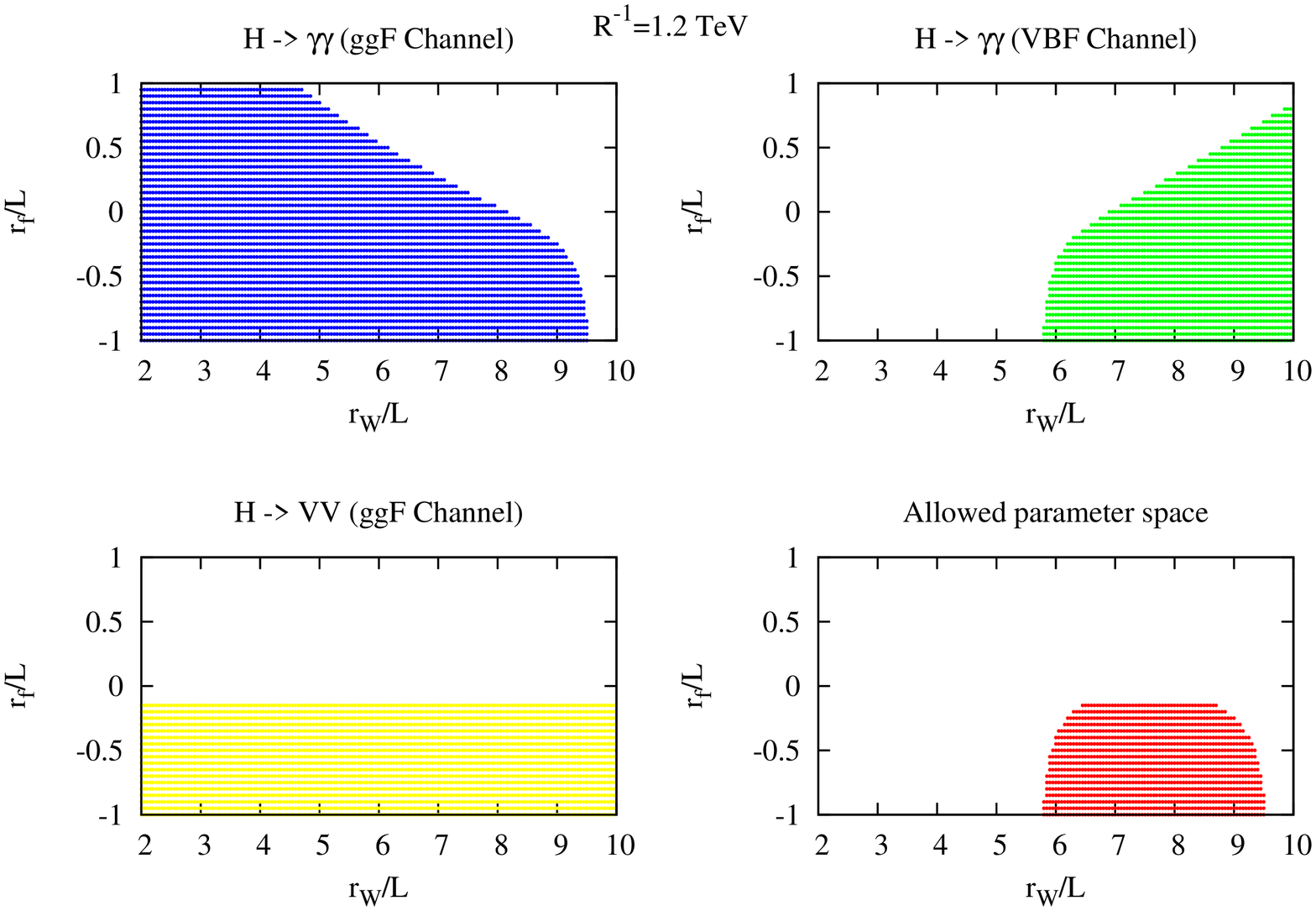}
\includegraphics[angle =-90, width=0.49\textwidth]{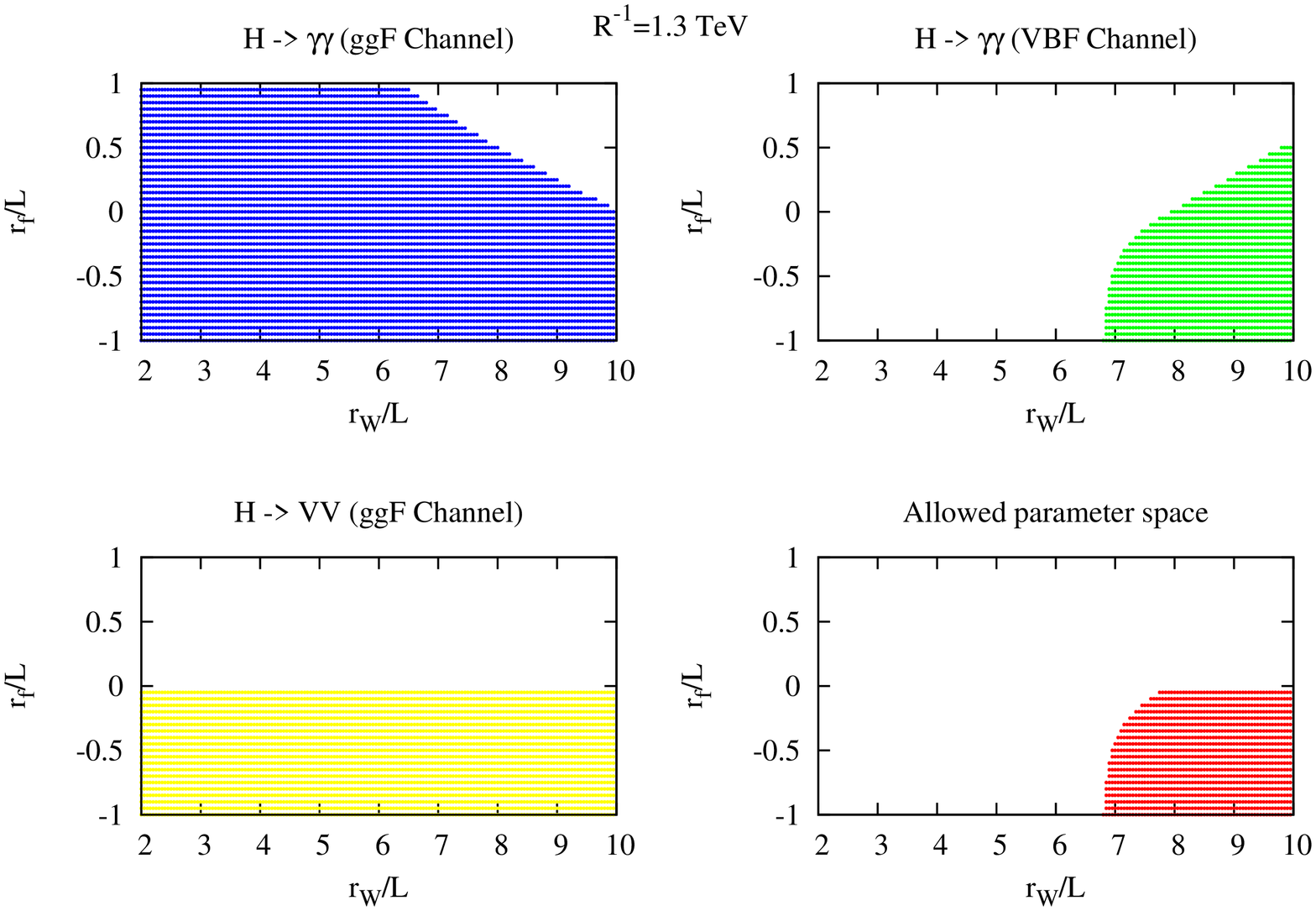}
\end{center}
\caption{Scattered points in $r_W/L$-$r_F/L$ plane which are consistent with the combined best-fit results of $\mu_{ggF}^{H\to \gamma\gamma}$, $\mu_{VBF}^{H\to \gamma\gamma}$, $\mu_{ggF}^{H\to VV}$ and all three together for $R^{-1}=1.2$ TeV (left panel) and $R^{-1}=1.3$ TeV (right panel).}
\label{nmued_higgs}
\end{figure}

In the framework of nmUED, the masses of KK-particles depends on $R^{-1}$ as well as on the coefficients ($r_G,~r_W,~r_B,~r_F$ e.t.c.) of the boundary localized kinetic terms. Therefore, the quasi-degeneracy between KK-$W$ and KK-top quark in the mUED model can be removed by suitable choice of the boundary parameters for the $W$-boson ($r_W$) and top quark ($r_F$)\footnote{We have considered same boundary parameters for all fermions flavors and thus, we have used the symbol $r_F$ instead of $r_t$. However, the bounds derived in the section are applicable only on the boundary parameters for the top quark. Constraints on the boundary parameters for the other quarks come from the direct collider searches for the KK-particles which will be discussed in the next section.}. For example, Fig.~\ref{mass} shows that for a particular value of $R^{-1}$, large $r_W$ and small $r_F$  give rise to a huge splitting between level-1 KK-$W$ and KK-top quark masses with  KK-top quark mass being larger than the  KK $W$-boson mass. In this scenario (large $r_W$ and small $r_F$), it is possible to enhance the Higgs to di-photon decay branching ratio.  In our analysis, we have scanned the relevant part of nmUED parameter space (in particular scanned over $r_W$ and $r_F$) for several fixed values of $R^{-1}$ and computed different Higgs signal strengths. Then we have checked the consistency of these signal strengths with the combined best-fit values in Table~\ref{bfss} within its error bars. Our results are presented in Fig.~\ref{nmued_higgs} for $R^{-1}=1.2$ TeV (left panel) and $R^{-1}=1.3$ TeV (right panel). In Fig.~\ref{nmued_higgs}, we have presented the scattered points in $r_W/L$-$r_F/L$ plane which are consistent with the combined best-fit results of $\mu_{ggF}^{H\to \gamma\gamma}$ (top left panel), $\mu_{VBF}^{H\to \gamma\gamma}$ (top right panel) and $\mu_{ggF}^{H\to VV}$ (bottom left panel). In the bottom right panel of Fig.~\ref{nmued_higgs}, we have presented the points which are consistent with all combined best-fit values of the measured Higgs signal strengths. Fig.~\ref{nmued_higgs} shows that in the framework of nmUED, large positive $r_W$ and  negative $r_F$ can explain all the measured Higgs signal strengths.
\begin{table}[h]
\begin{center}
\begin{tabular}{||c||c|c||c|c||c|c||c||c|c|c||}
\hline \hline
Cuts & \multicolumn{2}{|c||}{A (2-jets)} & \multicolumn{2}{|c||}{B (3-jets)} & \multicolumn{2}{|c||}{C (4-jets)} & D  & \multicolumn{3}{|c||}{E (6-jets)} \\\cline{2-11}
 & L & M & M & T & M & T & (5-jets)  & L & M & T \\\hline\hline 
$E_T\!\!\!\!\!\!/~>$ [GeV] & \multicolumn{10}{|c||}{160}\\\hline
$p_T^{j_1}>$ [GeV] & \multicolumn{10}{|c||}{130}\\\hline
$p_T^{j_2}>$ [GeV] & \multicolumn{10}{|c||}{60}\\\hline
$p_T^{j_3}>$ [GeV] & \multicolumn{2}{|c||}{-} & \multicolumn{2}{|c||}{60}  & \multicolumn{2}{|c|}{60} & 60 & \multicolumn{3}{|c||}{60} \\\hline 
$p_T^{j_4}>$ [GeV] &  \multicolumn{2}{|c||}{-} & \multicolumn{2}{|c||}{-}  & \multicolumn{2}{|c|}{60} & 60 & \multicolumn{3}{|c||}{60} \\\hline
$p_T^{j_5}>$ [GeV] &  \multicolumn{2}{|c||}{-} & \multicolumn{2}{|c||}{-}  & \multicolumn{2}{|c|}{-} & 60 & \multicolumn{3}{|c||}{60} \\\hline
$p_T^{j_6}>$ [GeV] &  \multicolumn{2}{|c||}{-} & \multicolumn{2}{|c||}{-}  & \multicolumn{2}{|c|}{-} & - & \multicolumn{3}{|c||}{60} \\\hline
$\Delta \phi(j_i,\vec E_T\!\!\!\!\!\!/~)_{min}>$ & \multicolumn{4}{|c||}{0.4 \{i=1,2,3 if $p_T^{j_3}>40$ GeV\}}& \multicolumn{6}{|c||}{0.4 \{i=1,2,3\}, 0.2 $p_T^{j_i}>40$ GeV}\\\hline
$E_T\!\!\!\!\!\!/~/M_{eff}(N_j)>$ & 0.2 & - & 0.3& 0.4& 0.25& 0.25& 0.2& 0.15& 0.2& 0.25\\\hline  
$m_{eff}(incl.)$ [TeV] & 1.0 & 1.6 & 1.8 & 2.2 & 1.2 & 2.2 & 1.6 & 1.0 & 1.2 & 1.5 \\\hline\hline
$\sigma_{BSM}$ [fb] & 66.07 & 2.52 & 0.73 & 0.33 & 4.00 & 0.12 & 0.77 & 4.55 & 1.41 & 0.41 \\\hline\hline

\end{tabular}
\end{center}
\caption{Cuts used by the ATLAS collaboration to define the signal regions. $\Delta \phi(jet,\vec E_T\!\!\!\!\!\!/~)$ is the azimuthal separations between $\vec E_T\!\!\!\!\!/~$ and the reconstructed jets. ${m_{eff}(N_j)}$ is defined to be the scalar sum of the transverse momenta of the leading $N$ jets together with $E_T\!\!\!\!\!\!/~$. However, for $m_{eff}^{incl.}$, the sum goes over all jets with $p_T>40$ GeV. Last column corresponds to the 95\% C.L. observed  upper limits on the non-SM contributions $\sigma_{BSM}$.}
\label{ATLAS_limits}
\end{table}

\begin{table}[h]
\begin{center}
\begin{tabular}{||c|c|c|c|c|c||}
\hline \hline
\multicolumn{6}{||c||}{Benchmark Point (BP)}\\
\hline\hline
$R^{-1}$ & $\Lambda R$ & $\mu L$ & $r_g/L$ & $r_F/L$ & $r_W/L$ \\\hline
1.2 TeV & 5    & -0.02  & -0.05 & -0.42 & 7.4 \\\hline\hline
\multicolumn{6}{||c|}{Masses in GeV}\\
\hline\hline
$m_{Q^{(1)}}$ &$m_{L^{(1)}}$ &$m_{G^{(1)}}$ &$m_{W^{(1)\pm}}$ &$m_{Z^{(1)}}$ &$m_{\gamma^{(1)}}$ \\\hline
1800 & 1800 & 1265 & 275 & 275 & 260 \\\hline\hline 
\end{tabular}
\end{center}
\caption{Benchmark Point}

\label{BP}
\end{table}
\subsection{Collider Phenomenology}
 The KK-quarks and gluons carry colors charges and the LHC is a proton-proton collider. Therefore, it is needless to mention that the production cross sections of TeV scale level-1 KK-quarks ($q^1$) and KK-gluons ($g^1$) are large at the LHC. However, due to the conservation of KK-parity, the level-1 particles can only be pair produced and the pair production takes place via the tree level KK-number conserving interactions only. After being produced, the level-1 KK-quarks and gluons decay into lighter KK-particles in association with one or more SM particles. The former decay, in turn, producing more stable SM particles. Finally, the decay cascade terminates at the production of LKP. LKP, being stable and weakly interacting, escapes the detector without being detected and gives rise to missing energy signature. The pair production followed by the decay of level-1 KK-quarks and/or gluons in the framework of KK-parity conserving UED scenarios gives rise to multiple jets, leptons and missing transverse energy signature at the collider experiments.  Therefore, the collider phenomenology of KK-parity conserving UED models is quite similar to the $R$-parity conserving supersymmetry scenarios in which the pair production of squarks and/or gluinos gives rise to multiple jets, leptons and missing transverse energy in the final state. In $R$-parity conserving supersymmetry, missing transverse energy results from the lightest supersymmetric particle which is weakly interacting and stable as a consequence of $R$-parity conservation. The CMS \cite{Chatrchyan:2013lya} and ATLAS {\cite{ATLAS-CONF-2013-047} collaborations have searched for SUSY in jets + leptons + $E_T\!\!\!\!\!\!/~$ channel and in the absence of significant excess of signal events over the SM backgrounds, they put stringent bounds on the masses of squarks and the gluino in the framework of cMSSM using 7/8 TeV data. For example, with integrated luminosity of 20.3 fb$^{-1}$, in the context of constrained minimal supersymmetric SM scenario, equal masses of squarks and gluino are excluded below 1.7 TeV in the  jets + 0$l$ +$E_T\!\!\!\!\!\!/~$ channel from 8 TeV LHC data \cite{ATLAS-CONF-2013-047}. However, there are no such bounds on the masses of level-1 KK-quarks and gluons are available from the ATLAS or CMS collaborations in the context of UED scenarios. In this section, we have used ATLAS {\cite{ATLAS-CONF-2013-047} results in order to set bounds on the masses of level-1 KK-quarks and gluons and constrain the parameter space of nmUED. 

\begin{table}[h]
\begin{center}
\begin{tabular}{||c|c|c|c|c|c||}
\hline \hline
\multicolumn{6}{||c||}{Benchmark Point (BP)}\\
\hline\hline
$R^{-1}$ & $\Lambda R$ & $\mu L$ & $r_g/L$ & $r_F/L$ & $r_W/L$ \\\hline
1.2 TeV & 5    & -0.02  & -0.05 & -0.42 & 7.4 \\\hline\hline
\multicolumn{6}{||c|}{Masses in GeV}\\
\hline\hline
$m_{Q^{(1)}}$ &$m_{L^{(1)}}$ &$m_{G^{(1)}}$ &$m_{W^{(1)\pm}}$ &$m_{Z^{(1)}}$ &$m_{\gamma^{(1)}}$ \\\hline
1800 & 1800 & 1265 & 275 & 275 & 260 \\\hline\hline 
\end{tabular}
\end{center}
\caption{Benchmark Point}

\label{BP}
\end{table}

Before going into the details of our analysis, let us briefly introduce the multijet search \cite{ATLAS-CONF-2013-047} strategies and results of the ATLAS collaboration. A search for 2-6 jets in association with large $E_T\!\!\!\!\!\!/~$ at the LHC with $\sqrt s=8$ TeV and 20.3 $ fb^{-1} $ integrated luminosity has been communicated by the ATLAS collaboration. Jet candidates are reconstructed using the anti-kt jet clustering algorithm~\cite{anitkt} with a distance parameter of 0.4 in the rapidity coverage $|\eta|\le 4.9$. Electron (muon) candidates are required to have $p_T > 20(10)$ GeV and $|\eta| < 2.47(2.4)$. After identifying jets and lepton, any jet candidate lying within a distance $\Delta R = \sqrt{\Delta \eta^2 + \Delta \phi^2} < 0.2$ of an electron is discarded. A lepton candidate is removed from the list is it falls with in a distance $\Delta R=0.4$ of any survived jet candidate. Missing transverse momentum is reconstructed using all remaining jets and leptons and all calorimeter clusters not associated to such objects. Finally, jets with $|\eta|>2.8$ are removed from the list. After the object reconstruction, events with zero electron (muon) with $p_T>20(10)$ GeV are selected for further analysis. ATLAS collaboration has presented results for five inclusive analysis channels, characterized by increasing jet multiplicity from 2 to 6. In Table~\ref{ATLAS_limits}, we have presented the cuts used by the ATLAS collaboration to define the signal regions. For all the signal regions (SRs) defined by the ATLAS collaboration, good agreement is seen between the numbers of events observed in the data and the numbers of events expected from SM processes. As a result, 95\% C.L. upper limits are set on the  beyond SM cross-sections ($\sigma_{BSM}$) for different SRs.

\begin{table}[h]
\begin{center}
\begin{tabular}{||c||c|c||c||}
\hline \hline
Process & \multicolumn{2}{c||}{Supersymmetry} & nmUED \\
        & \multicolumn{2}{c||}{$\tilde g \tilde g$ one-step} & $g^{(1)}g^{(1)}$ one-step \\\hline
Point & \multicolumn{2}{c||}{$m_{\tilde g}=1265$ GeV} &  $m_{g^{(1)}}=1265$ GeV \\
      & \multicolumn{2}{c||}{$m_{\tilde \chi^\pm_1}=865$ GeV} & $m_{\tilde W^{(1)\pm}}=865$ GeV \\
      & \multicolumn{2}{c||}{$m_{\tilde \chi^0_1}=465$ GeV} & $m_{\tilde \gamma^{(1)}}=465$ GeV \\\hline
Cuts  &  \multicolumn{3}{c||}{Absolute efficiency in \%}\\\cline{2-4}
 (E-tight)  & ATLAS & Our Simulation & Our Simulation \\
& Appendix-C of \cite{ATLAS-CONF-2013-047} & & \\\hline\hline
$0$-lepton & 63.5 & 66.1& 57.3\\
$E_T\!\!\!\!\!/~>160$ GeV & 55.6 & 57.6& 54.7\\
$p_T^{j_1}>130$ GeV &55.6 & 57.5& 54.7\\
$p_T^{j_2}>60$ GeV & 55.6 & 57.5& 54.6\\
$p_T^{j_3}>60$ GeV & 55.4 & 57.3& 51.8\\
$p_T^{j_4}>60$ GeV & 53.4 & 55.2& 41.3\\
$p_T^{j_5}>60$ GeV & 46.3  & 47.1& 27.4\\
$p_T^{j_6}>60$ GeV &  31.7 & 31.1& 15.0\\
$\Delta \phi(j_{i},E_T\!\!\!\!\!/~),i=1,2,3$ & 26.5 & 26.1&12.2\\
$\Delta \phi(j,E_T\!\!\!\!\!/~),p_T^j>40$ GeV & 21.3 & 21.6&9.7\\
$E_T\!\!\!\!\!/~/m_{eff}(N_j)>$ 0.25 & 12.0 & 12.7 & 4.7\\
$m_{eff}(incl.)>1.5$ TeV & 7.9 & 8.3 & 4.5\\\hline\hline
\end{tabular}
\end{center}
\caption{Cut-flow table.}

\label{cuts_flow}
\end{table}

\begin{figure}[t]
\begin{center}
\includegraphics[angle =-90, width=1.0\textwidth]{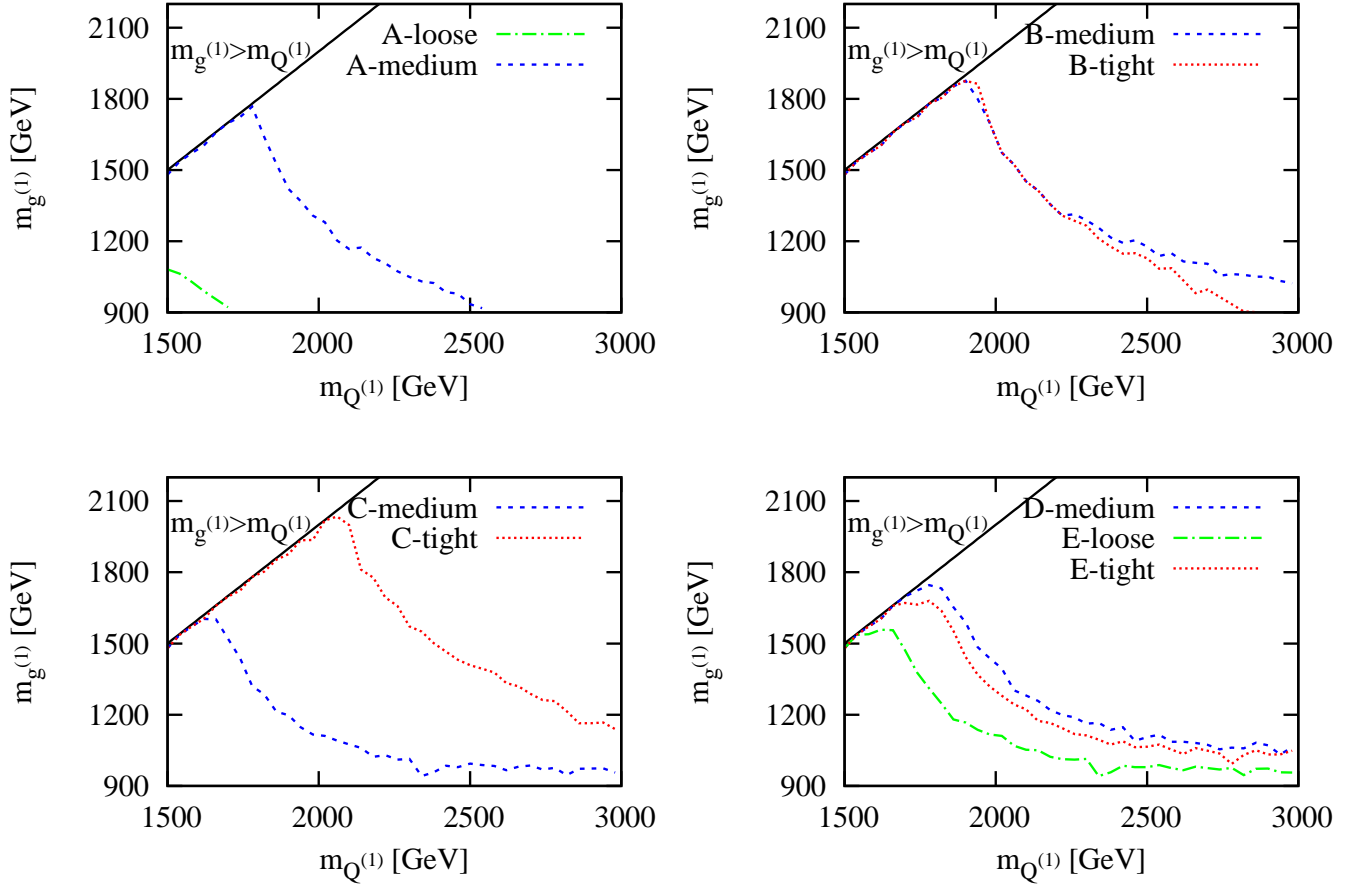}
\end{center}
\caption{The exclusion limits on $m_{Q^{1}}$-$m_{g^{1}}$ plane from 8 TeV 20.3 inverse femtobarn integrated luminosity ATLAS data for different ATLAS defined signal regions. We have assumed fixed mass for the level-1 electroweak KK gauge bosons ($m_W^{(1)\pm}=m_{Z^{(1)}}=275$ GeV and $m_W\gamma^{(1)}=260$ GeV).}
\label{mass_bound}
\end{figure}

After discussing the ATLAS SUSY multijets + $E_T\!\!\!\!\!\!/~$ search strategies  and results, we are now equipped enough to apply these results for nmUED scenario. We have used {\bf PYTHIA} \cite{pythia} for generating parton level $q^1q^1$, $g^1g^1$ and $q^1g^1$ events as well as for simulating the level-1 KK-quarks and gluon decay, ISR, FSR, hadronization e.t.c. However, the nmUED mass spectrum and decay branching fractions are calculated in CalcHEP 2.5 \cite{calchep} and then passed on to PYTHIA via the SUSY/BSM Les Houches Accord (SLHA) (v1.13) \cite{slha}. In our analysis, we have introduce a set of basic selection criteria to identify electrons, muons, jets and missing transverse energy. We have closely followed ATLAS collaborations suggested object reconstruction criteria and cuts described in the previous paragraph and Table~\ref{ATLAS_limits}.  For presenting our numerical results,  we have chosen a benchmark point (BP), listed in Table~\ref{BP}. Before going into the discussion of collider bounds on the masses of KK-quarks and gluons on the basis of ATLAS SUSY multijets + $E_T\!\!\!\!\!\!/~$ results, it is important to show the consistency of our analysis with the analysis done by the ATLAS collaboration \cite{ATLAS-CONF-2013-047}. To check the consistency of our analysis with the ATLAS analysis, we have considered the cut-flow table (Table 5) in Appendix C of Ref.~\cite{ATLAS-CONF-2013-047}. We have simulated gluino-gluino production followed by a one-step decay into neutralino using {\bf PYTHIA}.  Then we have analyzed those gluino-gluino events using our analysis code and presented the cut-flow chart in Table~\ref{cuts_flow} (3rd column). We have also presented the corresponding ATLAS simulated cut-flow chart (from Appendix C of Ref.~\cite{ATLAS-CONF-2013-047}) in the 2nd column of Table~\ref{cuts_flow}. Second and third column of Table~\ref{cuts_flow} shows that our simulation is in good agreement with the ATLAS simulation. In Table~\ref{cuts_flow}, we have also presented the cut-flow table for nmUED BP.

Our final results are presented in Fig.~\ref{mass_bound}. Assuming fixed mass for the level-1 electroweak KK gauge bosons ($m_W^{(1)\pm}=m_{Z^{(1)}}=275$ GeV and $m_W\gamma^{(1)}=260$ GeV), we have scaned the masses for the KK-quarks and gluons over a range between 800 GeV to 3 TeV. The exclusion limits on $m_{Q^{1}}$-$m_{g^{1}}$ plane for different ATLAS defined signal regions are presented in Fig.~\ref{mass_bound}. Fig.~\ref{mass_bound} shows that in the framework of nmUED, the strongest bound on the masses of level-1 KK-quarks and gluons arise from the C-tight signal region and equal mass level-1 KK-quark and gluon is ruled out below about 2.1 TeV. The mass exclusion results presented in Fig.~\ref{mass_bound} can be translated to put bounds on the boundary parameters for the gluon and quark in the framework of nmUED. In Fig.~\ref{param}, we have presented the allowed parameter space in $r_f/L$-$r_g/L$ plane after LHC 8 TeV run for two different values of $R^{-1}=$ 1.2 (left panel) and 1.3 TeV (right panel). 

\begin{figure}[t]
\begin{center}
\includegraphics[angle =-90, width=1.0\textwidth]{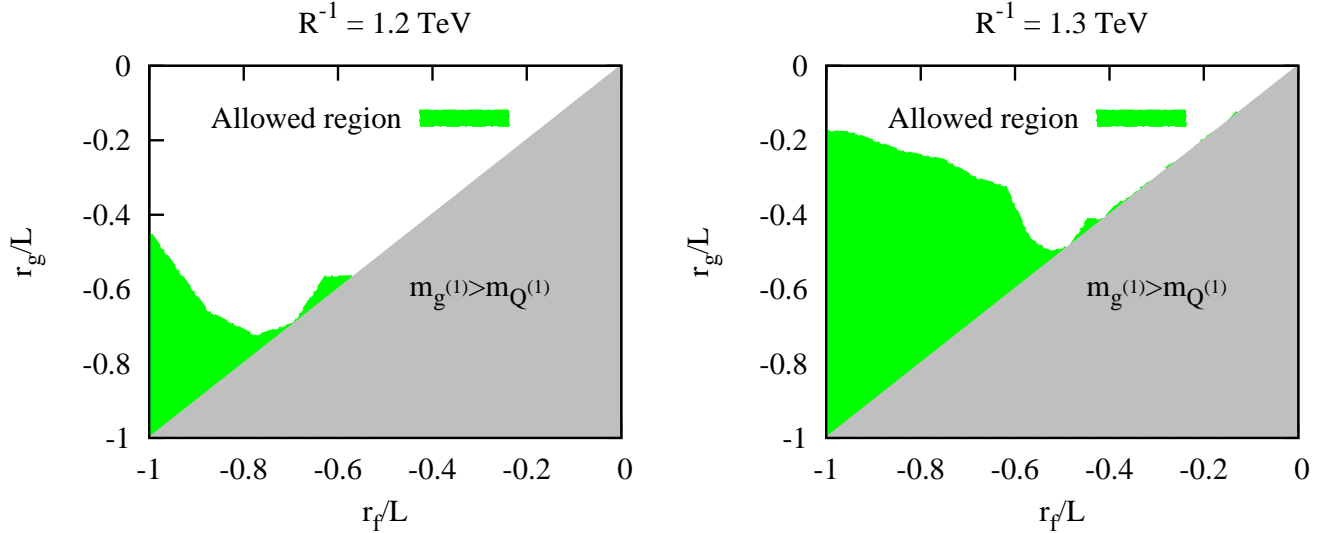}
\end{center}
\caption{Allowed parameter space in $r_f/L$-$r_g/L$ plane after LHC 8 TeV data.}
\label{param}
\end{figure}

\section{Summary}

{ We have investigated the phenomenology of the non-minimal Universal Extra-dimension model in the context of LHC Higgs results and ATLAS multijets + missing energy data.  We find that the minimal Universal extra dimension model (mUED) can not explain the the LHC Higgs data. However, the non-minimal Universal Extra Dimension (nmUED) with  suitable choice of boundary localized kinetic terms (BLKT) for 5-dimensional fermions and gauge bosons is consistent with all the LHC Higgs results.  The measured Higgs signal strengths in different production and decay channel put stringent bounds on the boundary parameters for the $SU(2)_W$ gauge bosons and quarks (in particular top quark). The enhanced  Higgs to di-photon decay rate, hinted by the ATLAS collaboration, points towards a large splitting between level-1 $W$-boson and top quark mass with level-1 top being heavier than the  $W$-boson. The bound on the masses of level-1 gluon and light quarks arises from the multijets + missing energy searches at the LHC experiment. We have used ATLAS multijets + missing energy analysis and computed bounds on the masses of level-1 KK gluon and quarks. The strongest lower bound is obtained for equal mass level-1 quark and gluon masses (for the BLKT parameters satisfying all the Higgs data) and is $2.1$ TeV}.
 
{Another interesting phenomena we found is regarding the topological structure of the  final state consisting of high $p_T$ multijets plus large missing energy.  The collider phenomenology of models with Universal Extra Dimensions (UED) is well know to be  similar to that of supersymmetric (SUSY) scenarios with a compressed SUSY spectrum, and for that reason, mUED has been called bosonic supersymmetry. 
However, in nmUED, with the suitable choice of parameters to explain the LHC Higgs data,  we find that BLK terms remove the degeneracy in the KK mass spectrum and thus, pair production of level-1 quarks and gluons at the LHC gives rise to hard jets, leptons and large missing energy in the final state. Thus the topological structure of the final states is very similar to the usual(non-compressed) SUSY. Of course, the cross sections of these final states will be different for the nmUED and the SUSY case because of the different spins of the produced parent particles. But LHC will not be able to to distinguish it because we do not know the masses of these particle. Thus the production of the 2nd excites in nmUED will be the key to distinguish it from the supersymmetry at the LHC}.\\
{\bf Acknowledgment:}  The work of KG, DK and SN was supported in part by the U.S. Department of Energy
Grant Number DE-SC0010108.

\end{document}